\documentclass[reprint,superscriptaddress,amsmath,amsfonts,amssymb,aps,prb,showkeys]{revtex4-2}

\usepackage{hyperref}
\usepackage{graphicx}
\usepackage{mathrsfs}

\hypersetup{
  breaklinks = {true},
  citecolor = {blue},
  colorlinks = {true},
  linkcolor = {red},
}

\usepackage{physics}
\usepackage{mathtools}
\usepackage[dvipsnames]{xcolor}
\usepackage{braket} 
\usepackage{upgreek}

\usepackage{ulem}

\begin{document}

\title{Scaling of the Integrated Quantum Metric in Disordered Topological Phases}

\author{Jorge Martínez Romeral}
\affiliation{Catalan Institute of Nanoscience and Nanotechnology (ICN2), CSIC and BIST, Campus UAB, Bellaterra, 08193 Barcelona, Spain}
\affiliation{Department of Physics, Campus UAB, Bellaterra, 08193 Barcelona, Spain}
\author{Aron W. Cummings}
\affiliation{Catalan Institute of Nanoscience and Nanotechnology (ICN2), CSIC and BIST, Campus UAB, Bellaterra, 08193 Barcelona, Spain}

\author{Stephan Roche}
\affiliation{Catalan Institute of Nanoscience and Nanotechnology (ICN2), CSIC and BIST,
Campus UAB, Bellaterra, 08193 Barcelona, Spain}
\affiliation{ICREA--Instituci\'o Catalana de Recerca i Estudis Avan\c{c}ats, 08010 Barcelona, Spain}
\date{\today}

\begin{abstract}
We report a study of a disorder-dependent real-space representation of the quantum geometry in topological systems. Thanks to the development of an efficient linear-scaling numerical methodology based on the kernel polynomial method, we can explore nontrivial behavior of the integrated quantum metric and Chern number in disordered systems with sizes reaching the experimental scale. We illustrate this approach in the disordered Haldane model, examining the impact of Anderson disorder and vacancies on the trivial and topological phases captured by this model.
\end{abstract}

\maketitle

\section{Introduction}
During the last decades, the Berry curvature has been the cornerstone of topological matter \cite{KOHMOTO1985343, RevModPhys.82.1959}. Recently, it has been shown that it is only one part, the imaginary part, of a general structure called the quantum geometric tensor (QGT). The real part of the QGT, called the quantum metric, has received a lot of recent attention for its role in flat-band superconductivity \cite{FirstPavi, AnotherPavi, boundedMATBGBernevig, reviewSCandSFinflatbands, GuineaMultilayerspublished,Rossi2021Oct,EvidenceFBSCbyQGT,MinimalcouplingQM}, electron-phonon coupling \cite{electronphononcoupling}, linear \cite{Queiroz1, Queiroz2, RaquelQueirozStepResponse, quantumtransportindispernsionless} and nonlinear response theory \cite{UnificationQAHE, topologicalmagnetnonlinearQM, Elotrodetopologicalmagnet, intrinsicnonlinearsigma,nonlinear1,Positionalshift,UnificationofNHEbyQGT}, and a compendium of other effects \cite{FractionalChernInsulatorsInWalgrebra,QFHEinTI,StabilityofFHE,EssayTorma,ReviewQGTSept}. The quantum metric is a key quantity in the modern theory of insulators, as it is directly related to the localization of the ground state \cite{insulatingtheoryofmatter, localtheoryoftheinsulating, MaximallylocalizedWannierfunctions, PedroGuerrero2024Jan}. It has also proven to be useful for the estimation of topological quantities, as its trace is bounded from below by the absolute value of the Berry curvature  \cite{FirstPavi, bandgeometryoffractionalTI}. In addition, the QGT has already been experimentally investigated and its connection to various observables has been demonstrated \cite{ExperimentalSCqubit, WGTmanifold, Experimentaldiamonds, ExperimentalRaman, ExperimentalHalldrift, topologicalmagnetnonlinearQM}. We note that the single-particle quantum metric has been connected to important features in the strongly interacting regime \cite{electronphononcoupling, reviewSCandSFinflatbands, GuineaMultilayerspublished}.

Although the Berry curvature has been widely studied in the presence of disorder \cite{JoseHugoKPM, ComputationProjectorKPM, VacanciesChernInsulator}, the scaling properties of the quantum metric in complex disordered models remains mostly unexplored. Since disorder induces localization in low-dimensional systems \cite{Anderson1958Mar, 50yearsoftheAndersonmodel}, and given that the quantum metric is a direct measure of the localization of the ground state \cite{insulatingtheoryofmatter}, the behavior of the disorder-dependent quantum metric is expected to provide relevant information for understanding transport properties. However, such studies are scarce \cite{marsal2024enhanced, PedroGuerrero2024Jan, Onedimensionaldisorder}, and little is known for disordered topological phases. 

In this Letter, we develop a computationally efficient real-space approach enabling the calculation of the  integrated quantum geometric tensor (IQGT), i.e., the integral of the quantum geometric tensor over the first Brillouin zone. This approach allows us to investigate large, arbitrarily disordered systems. We then apply this method to the study of the Haldane Hamiltonian \cite{Haldane1988Oct, Donna2006} in the presence of various sources of disorder. By tuning the Hamiltonian's parameters, we study the impact of disorder on both the nontrivial Chern insulating phase and the topologically trivial phase. Disorder is introduced either via the random Anderson potential \cite{Anderson1958Mar}, or through a random distribution of vacancies.

To account for the disorder-induced breaking of translational invariance and to simulate large systems, we present a real-space linear-scaling approach based on the kernel polynomial method \cite{LSQUANT}. This method is similar to implementations of the real-space Chern marker \cite{TopologicalMarkerResta, ComputationProjectorKPM,Assuncaotopomarker,GrushinTopomarkeramorphous,Prodan2010,LocalTopologicalMarkersinOddSpatialDimensionsandTheirApplicationtoAmorphousTopologicalMatter,Theoryoflocaltopologicalmarkersorfiniteandperiodictwo-dimensionalsystems}, and can handle systems containing many millions of atoms, allowing the study of materials on length scales relevant to experiments. In contrast to previous methods, our approach also allows \textit{both open and periodic boundary conditions}. These features are important even in large-area disordered samples, as edge and finite-size effects can lead to spurious results when calculating bulk topological properties \cite{insulatingtheoryofmatter}. Our method also gives access to the spatially-resolved IQGT, hence informing about its real space fluctuations.

Our findings reveal that the IQGT, and in particular the integrated quantum metric, can display nontrivial behavior in the presence of disorder, exhibiting localization or delocalization depending on the disorder strength and topological phase. This is also displayed in the real-space projection of the integrated quantum metric, which offers more precise understanding of the quantity. Here we have focused on the well-known Haldane model to make meaningful comparisons between the scaling of the integrated quantum metric and already understood transport properties in the presence of disorder. But looking ahead, this method is completely general to any single-particle Hamiltonian and thus may be used explore a variety of other types of materials for which disorder effects could be key in understanding their emergent properties.

\section{The quantum geometric tensor}
In single-particle periodic systems, the QGT can be written as a momentum-dependent quantity, 
\begin{gather}
\mathit{q}_{\alpha\beta}(\mathbf{k}) = \nonumber \\
\dfrac{1}{\pi} \sum_{ij} f_{i\mathbf{k}} (1-f_{j\mathbf{k}})
\frac{\bra{\psi_{i\mathbf{k}}} \partial_{k_\alpha}\hat{H_{\mathbf{k}}} \ket{\psi_{j\mathbf{k}}} \bra{\psi_{j\mathbf{k}}} \partial_{k_\beta}\hat{H_{\mathbf{k}}} \ket{\psi_{i\mathbf{k}}}} {\left(E_{i\mathbf{k}}-E_{j\mathbf{k}}\right)^2}, \label{eq:QGTinKspace}
\end{gather}
where $\hat{H}_{\mathbf{k}}$ is the Bloch Hamiltonian, $\ket{\psi_{i\mathbf{k}}}$ is the eigenstate of $\hat{H}_{\mathbf{k}}$ in band $i$ at momentum $\mathbf{k}$, $E_{i\mathbf{k}}$ is its corresponding eigenenergy, and $f_{i\mathbf{k}}$ is its occupation \cite{insulatingtheoryofmatter, OriginalpaperQM}.
The QGT can be separated into two parts,
\begin{equation}\label{eq:qgt_real_imag}
\mathit{q}_{\alpha\beta}(\mathbf{k}) = g_{\alpha\beta}(\mathbf{k}) + \mathrm{i}\Omega_{\alpha\beta}(\mathbf{k})/2,
\end{equation}
where the antisymmetric imaginary part, $\Omega_{\alpha\beta}(\mathbf{k})$, is the Berry curvature, and the real symmetric part, $g_{\alpha\beta}(\mathbf{k})$, is the quantum metric \cite{insulatingtheoryofmatter}.

By introducing the state projectors
\begin{align}\label{eq:projectork}
    \hat{P}(\mathbf{k}) &= \sum_i f_{i\mathbf{k}}\ket{\psi_{i\mathbf{k}}}\bra{\psi_{i\mathbf{k}}}, \\
    \hat{Q}(\mathbf{k}) &= \sum_i (1-f_{i\mathbf{k}})\ket{\psi_{i\mathbf{k}}}\bra{\psi_{i\mathbf{k}}},
\end{align}
and the position operator
\begin{equation}
\bra{\psi_{i\mathbf{k}}}\hat{r}_{\alpha}\ket{\psi_{j\mathbf{k}}}=i\frac{\bra{\psi_{i\mathbf{k}}} \partial_{k_\alpha}\hat{H_{\mathbf{k}}} \ket{\psi_{j\mathbf{k}}}}{E_{i\mathbf{k}}-E_{j\mathbf{k}}},
\end{equation}
the QGT can be written as
\begin{align}
    \mathit{q}_{\alpha\beta}(\mathbf{k}) &= -\dfrac{1}{\pi} \sum_i\bra{\psi_{i\mathbf{k}}}\hat{P}(\mathbf{k})\hat{r}_{\alpha}\hat{Q}(\mathbf{k})\hat{r}_{\beta}\ket{\psi_{i\mathbf{k}}} \nonumber \\
    &= -\dfrac{1}{\pi} \sum_i\bra{\psi_{i\mathbf{k}}} \hat{P}(\mathbf{k}) \left[\hat{r}_{\alpha},\hat{P}(\mathbf{k})\right] \left[\hat{r}_{\beta},\hat{P}(\mathbf{k})\right] \ket{\psi_{i\mathbf{k}}},
\end{align}
where the final equality holds in the zero-temperature limit when $\hat{P}\hat{P} = \hat{P}$, and follows from the cyclic property of the trace.

Next we can define the integrated quantum geometric tensor as the integral of $\mathit{q}_{\alpha\beta}(\mathbf{k})$ over the Brillouin zone,
\begin{equation}\label{eq:iqgt}
    \mathcal{Q}_{\alpha\beta} = \int_{\mathrm{BZ}} \mathit{q}_{\alpha\beta}(\mathbf{k}) d\mathbf{k}.
\end{equation}
 \\
Following Eqs.\ \eqref{eq:qgt_real_imag} and \eqref{eq:iqgt}, the real and imaginary parts of $\mathcal{Q}_{\alpha\beta}$ respectively yield what we call the integrated quantum metric (IQM), as well as the Chern number \cite{OriginalpaperQM, Thouless1982Aug, Kane2005Sep},
\begin{align}\label{eq:C}
  \mathcal{G}_{\alpha\beta} &= \frac{1}{2}\Re{\mathcal{Q}_{\alpha\beta}}, \\
  \label{eq:IQM}
  \mathcal{C} &= \epsilon^{\alpha\beta} \Im{\mathcal{Q}_{\beta\alpha}},
\end{align} 
where $\epsilon^{\alpha\beta}$ is the Levi-Civita symbol.


Since the QGT is a positive semi-definite matrix \cite{topologyandQM}, the Berry curvature and the quantum metric are related via the inequality $\Tr g_{\alpha\beta}(\mathbf{k}) \geq \left|\Omega_{\alpha\beta}(\mathbf{k}) \right|$ \cite{FirstPavi}, thus yielding the relation
\begin{equation}\label{eq:iqgt_inequality}
    \Tr\mathcal{G}_{\alpha\beta} \geq \mathcal{C}
\end{equation}
between the IQM and the Chern number.

The IQM is related to the invariant part of the spread of the maximally localized Wannier functions \cite{MaximallylocalizedWannierfunctions}, which in two dimensions is expressed as $\Omega_{\mathrm{I}}=A/2\pi \cdot \Tr\mathcal{G}_{\alpha\beta}$, with $A$ the system area.
The IQM is a keystone of the modern theory of insulators,
which states that the dimensionless quantity $\Tr \mathcal{G}_{\alpha\beta}$ diverges in the thermodynamic limit ($A \rightarrow \infty$) for metals and converges to a finite value for any kind of insulator \cite{insulatingtheoryofmatter, localtheoryoftheinsulating}.


The relation of the IQM with the Wannier spread indicates that the IQM is a measure of the localization of the ground state. A low IQM implies localized Wannier functions, while a higher IQM suggests more delocalized electronic states and extended position fluctuations, which in the thermodynamic limit may indicate metallic behavior.



\section{Linear-scaling calculation of the IQGT}
We  can efficiently calculate the IQGT in large-area disordered systems by using its gauge-invariant real-space form \cite{localtheoryoftheinsulating, resta2018Insulatingmatter, TopologicalMarkerResta, OrbitalMagnetizationasaLocalProperty},

\begin{equation}\label{eq:QGTconmutators}
\mathcal{Q}_{\alpha\beta} = -\dfrac{4\pi}{A} \mathrm{Tr}\left\{ \hat{P} \left[\hat{r}_{\alpha},\hat{P}\right] \left[\hat{r}_{\beta},\hat{P}\right] \right\},
\end{equation}
where $\hat{P}$ is the Brillouin-zone integral of the projector in Eq.\ \eqref{eq:projectork}, defined in the zero-temperature limit as $\hat{P} = \Theta(\hat{H}-E_{\mathrm{F}})$ with $\Theta$ the step function and $E_{\mathrm{F}}$ the Fermi energy. Here the trace is taken over lattice sites in real space, and it has been demonstrated that Eq.\ \eqref{eq:QGTconmutators} is equivalent to Eq.\ \eqref{eq:iqgt} in the limit of infinite system size \cite{resta2018Insulatingmatter, TopologicalMarkerResta, OrbitalMagnetizationasaLocalProperty}.

Regarding the choice of terminology, Eq.\ \eqref{eq:QGTconmutators} has previously been referred to with a variety of names, such as the ``quantum metric tensor'' \cite{localtheoryoftheinsulating}, or the ``quantum metric'' \cite{marsal2024enhanced}. Here we refer to it as the ``integrated quantum geometric tensor,'' and its real part as the ``integrated quantum metric,'' to highlight that this expression is the integral of the quantum geometric tensor of Eq.\ \eqref{eq:QGTinKspace} over the Brillouin zone.

In periodic systems, computing the IQGT via Eq.\ \eqref{eq:QGTinKspace} requires diagonalization of the Bloch Hamiltonian, which becomes computationally expensive for systems with large unit cells.
Additionally, in systems with open boundary conditions or disorder, where periodicity is broken, direct calculation of Eq.\ \eqref{eq:QGTconmutators} relies on brute-force diagonalization of huge real-space Hamiltonians, which significantly restricts the accessible system sizes. To avoid direct diagonalization, we employ the kernel polynomial method (KPM) by expanding the ground-state projector as a series of Chebyshev polynomials \cite{LSQUANT,suppmaterial},
\begin{equation}
    \hat{P}(\widetilde{H})=\sum_{m=0}^{\infty}g_m\mu_m\hat{T}_m(\widetilde{H}),
\end{equation}
where $g_m$ are the coefficients of the Jackson kernel \cite{KPM2}, $\mu_m$ are the moments of the expansion \cite{LSQUANT}, and $\hat{T}_m(\widetilde{H})$ is the Chebyshev polynomial of order $m$. Finally, $\widetilde{H}$ is the Hamiltonian normalized such that its energy spectrum lies in $[-1,1]$; $\widetilde{H} = ( \hat{H}-\Bar{E} )/\Delta E$, where $\Bar{E} = (E_{\mathrm{max}}+E_{\mathrm{min}})/2$, $\Delta E = (E_{\mathrm{max}}-E_{\mathrm{min}})/2$, and $E_{\mathrm{min}}$ and $E_{\mathrm{max}}$ are the minimum and maximum eigenvalues of $\hat{H}$, respectively. As the zero temperature ground-state projector is simply the step function, the moments $\mu_m$ have an exact analytical form \cite{ComputationProjectorKPM}.

Previous implementations of similar formulas, to calculate the so-called topological markers \cite{TopologicalMarkerResta, ComputationProjectorKPM, GrushinTopomarkeramorphous,Prodan2010,LocalTopologicalMarkersinOddSpatialDimensionsandTheirApplicationtoAmorphousTopologicalMatter,Theoryoflocaltopologicalmarkersorfiniteandperiodictwo-dimensionalsystems}, relied on open boundary conditions since the position operator is ill-defined for periodic boundary conditions. Here, by using the recurrence relation of the Chebyshev polynomials, the commutator
\begin{equation}
\left[\hat{r}_{\alpha},\hat{P}\left(\widetilde{H}\right)\right]=\sum_{m=0}^\infty g_m\mu_m\left[\hat{r}_{\alpha},\hat{T}_m\left(\widetilde{H}\right)\right]
\end{equation}
can be expressed in terms of the velocity operator $\hat{v}_\alpha = [\hat{r}_{\alpha},\hat{H}] / \mathrm{i} \hbar$, which is well defined independent of the boundary conditions and choice of origin. The commutator can then be calculated using the recursive series,
\begin{equation}\label{eq:recursiverelation}
    \begin{aligned}
        &\left[\hat{r}_{\alpha},\hat{T}_m\left(\widetilde{H}\right)\right]= 2\widetilde{V}_{\alpha}\,\hat{T}_{m-1}\left(\widetilde{H}\right) \\
        &+2\widetilde{H}\left[\hat{r}_{\alpha},\hat{T}_{m-1}\left(\widetilde{H}\right)\right]-\left[\hat{r}_{\alpha},\hat{T}_{m-2}\left(\widetilde{H}\right)\right],
    \end{aligned}   
\end{equation}
where we have used the identity $[\hat{r}_{\alpha},\widetilde{H}] = \mathrm{i}\hbar\widetilde{V}_\alpha$, with $\widetilde{V}_\alpha = \hat{V}_\alpha/\Delta E$ the renormalized velocity operator.

We efficiently evaluate the trace of Eq.\ \eqref{eq:QGTconmutators} using the stochastic trace approximation \cite{KPMoutsidefromgroup},
\begin{equation}
    \mathrm{Tr}\{\hat{A}\} \approx \frac{1}{R} \sum_{i=1}^R \bra{\phi_i} \hat{A} \ket{\phi_i},
\end{equation}
where $R$ is the number of complex random vectors $\ket{\phi_i}$. The error of this approximation, which manifests as numerical noise, scales as $1/\sqrt{RN}$ where $N$ is the number of atoms in the system. A small value of $R$ for a given $N$ can thus lead to apparent particle-hole asymmetries in what should be a symmetric system, but these are progressively reduced by increasing $R$.

\begin{figure}[tbh]
\includegraphics[width=1 \columnwidth]{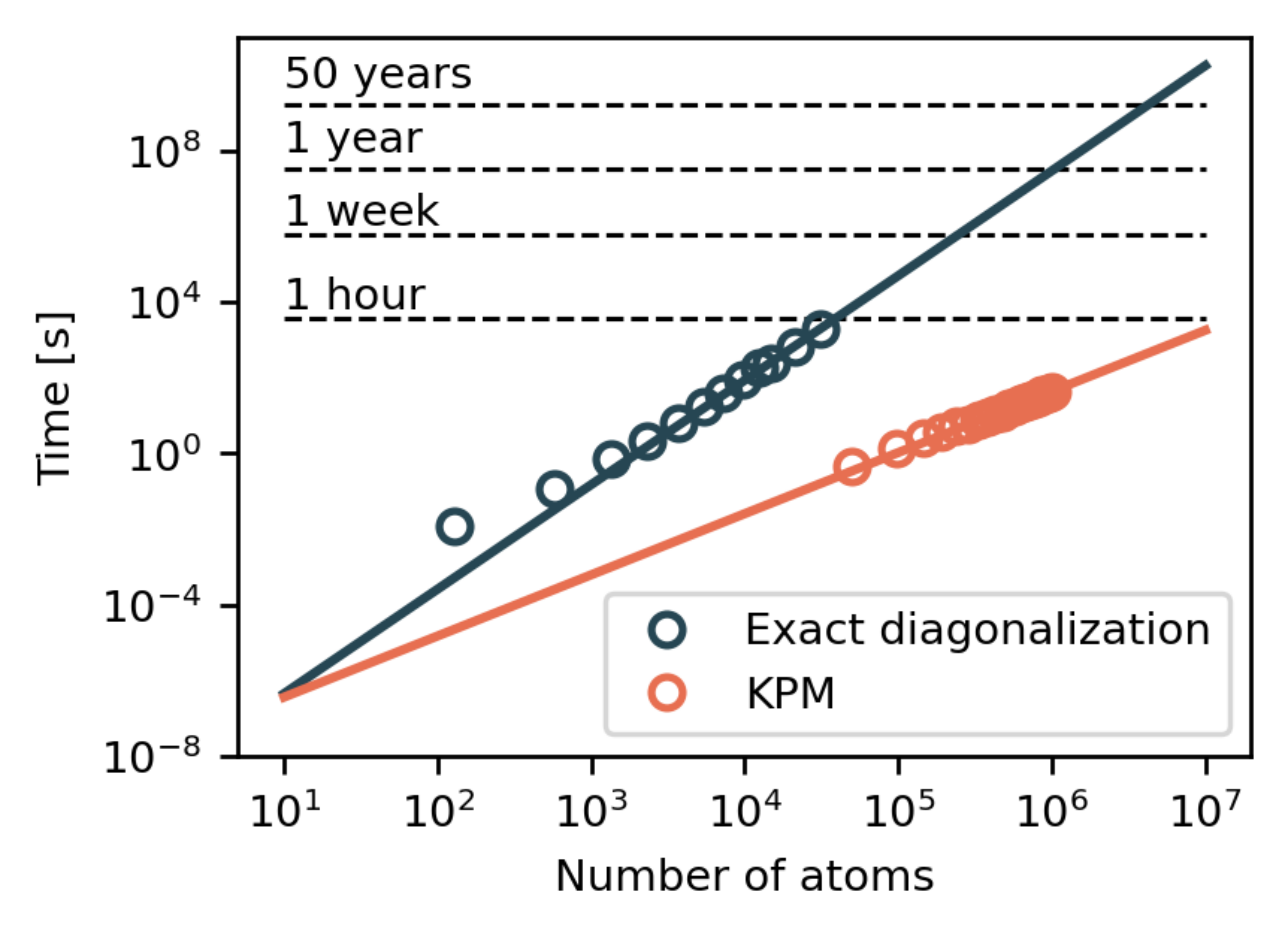}
\caption{\label{fig:comparison_with_time}Comparison of the computation time of our KPM method (orange) with exact diagonalization (blue), similar to the one done in \cite{marsal2024enhanced}. Symbols are directly measured computation times, and lines are extrapolations to larger systems. } 
\end{figure}

The combination of KPM and the stochastic trace approximation makes the method $\mathcal{O}(N)$, i.e., its computational cost scales linearly with the number of atoms or lattice sites $N$ \cite{LSQUANT, KPM2}. This is in contrast to direct diagonalization methods \cite{marsal2024enhanced}, which scale as $\mathcal{O}(N^2)$ or higher. A direct comparison can be seen in Fig.\ \ref{fig:comparison_with_time}, where the orange and blue symbols indicate the time it takes to compute the IQM in a sample of disordered graphene via our method and via exact diagonalization, respectively. The solid lines show the extrapolated computation time out to $N = 10^7$ atoms.

After calculating Eq.\ \eqref{eq:QGTconmutators}, the integrated quantum metric and the Chern number are given by Eqs.\ \eqref{eq:C} and \eqref{eq:IQM}. In practice, the Chebyshev polynomial expansion is performed for a finite number of polynomials $M$ and finite number of atoms $N$. We have verified that all calculations are converged with respect to both system size and $M$ \cite{suppmaterial}, with the exception of energies outside the band gap in the clean limit, where the IQM diverges as expected for metallic systems.

Equation \eqref{eq:QGTconmutators} is the trace over an operator in the real-space basis, and thus each element in the trace allows us to also construct a real-space map of the  IQGT. This local version of IQM, with the trace taken over a small number of sites, has been previously defined as a ``localization marker'' \cite{localtheoryoftheinsulating}, while the local imaginary part of the IQGT is typically referred to as a ``topological marker'' or ``local Chern marker/number'' \cite{TopologicalMarkerResta, ComputationProjectorKPM, GrushinTopomarkeramorphous,Prodan2010,LocalTopologicalMarkersinOddSpatialDimensionsandTheirApplicationtoAmorphousTopologicalMatter,Theoryoflocaltopologicalmarkersorfiniteandperiodictwo-dimensionalsystems}, and whose trace over the entire system yields the Chern number.

Here we briefly comment on the appearance of non-integer Chern numbers in the results below. Equation (2) yields the exact Chern number only in the thermodynamic limit, but one can never simulate a system with an infinite number of atoms. Thus, the appearance of seemingly non-integer Chern numbers is entirely due to finite-size effects, and one must perform a size-scaling analysis to extrapolate the Chern number to the thermodynamic limit \cite{Theoryoflocaltopologicalmarkersorfiniteandperiodictwo-dimensionalsystems, ComputationProjectorKPM, TopologicalmarkercurrentsinCherninsulators, LocalTopologicalMarkersinOddSpatialDimensionsandTheirApplicationtoAmorphousTopologicalMatter, Zhang2013}. We provide one such example in the SM \cite{suppmaterial}, indicating convergence towards integer values of $C$ with increasing system size, and a sharp topological transition with increasing disorder strength. Below we show results at a fixed system size with the understanding that while not in the thermodynamic limit, one may still reveal interesting behavior by examining the dependence of the IQM and $C$ on disorder strength. Meanwhile, we reiterate that all of our results for the IQM are converged with respect to system size, except for the case of clean systems outside the band gap, where the IQM diverges as expected for metallic systems.

\section{Application to the disordered Haldane model}
\begin{figure}[tbh]
\includegraphics[width=1\columnwidth]{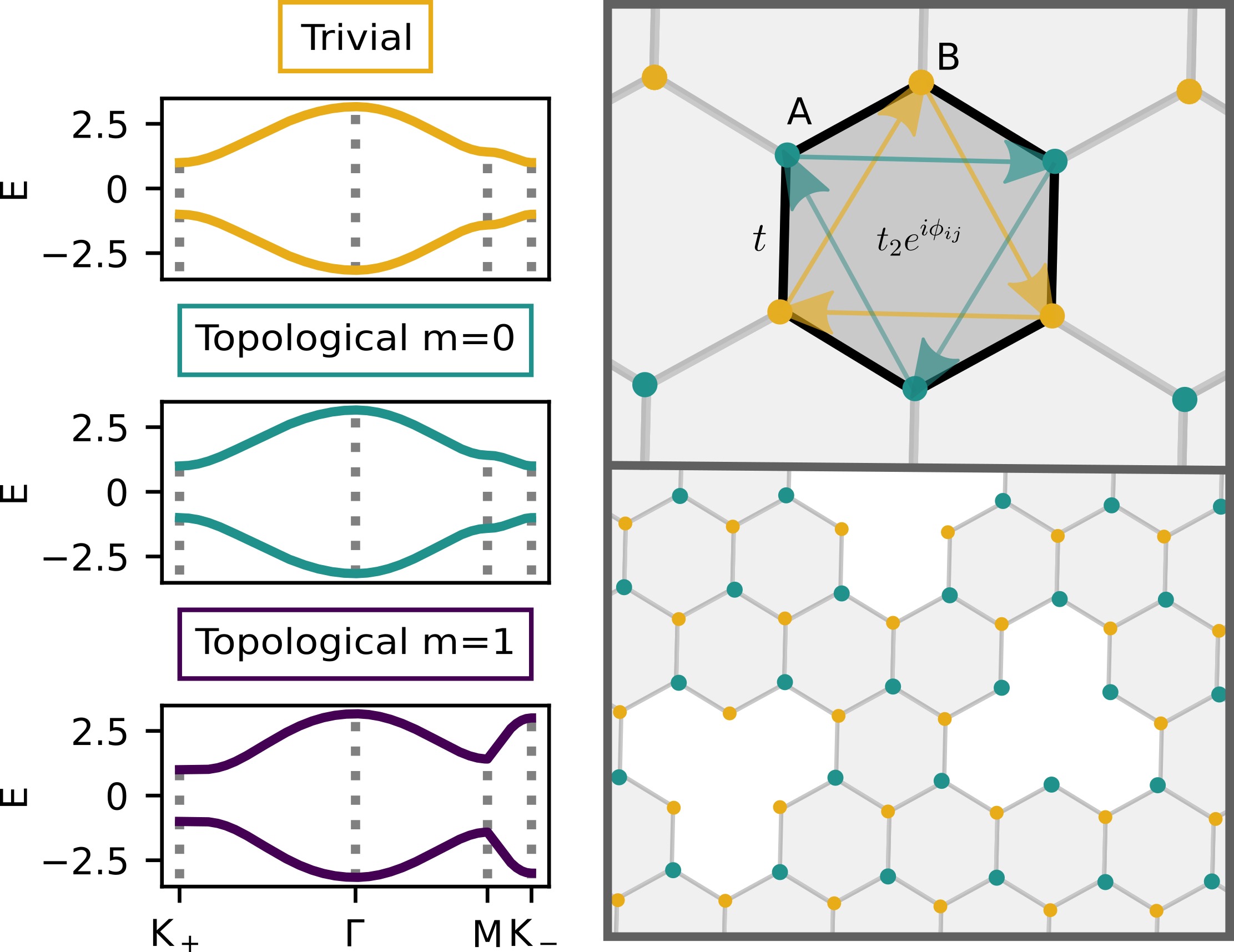}
\caption{\label{fig:Haldanechemea}Left panels: band structures of different phases of the Haldane model. Top left: trivial phase, $\Delta_m=1$ and $\Delta_{\mathrm{H}}=0$. Middle left: inversion-symmetric topological phase, $\Delta_m=0$ and $\Delta_{\mathrm{H}}=2$. Bottom left: topological phase with broken inversion symmetry $\Delta_m=2$ and $\Delta_{\mathrm{H}}=4$. Top right panel: spatial representation of the Haldane model where the arrows indicate complex second-nearest-neighbor hoppings and yellow/green dots indicate the A/B sublattices. Bottom right: real-space representation of the model with vacancies.}
\end{figure}

We now study the IQGT in the disordered Haldane model, a tight-binding model in a honeycomb lattice with one orbital per atom \cite{Haldane1988Oct}, shown schematically in Fig.\ \ref{fig:Haldanechemea} (top right panel). This model contains three terms,
\begin{equation}
    \hat{H} = t\sum_{\left<ij\right>}c_i^{\dagger}c_j\pm m\sum_{i\in A/B}c_i^{\dagger}c_i+t_2\sum_{\left<\left<ij\right>\right>}e^{i\phi_{ij}}c_i^{\dagger}c_j+\mathrm{h.c.}
\end{equation}
The first term is nearest-neighbor hopping, where we arbitrarily set $t=1$. All other energy scales are thus relative to $t$. The second term, positive (negative) on sublattice $A$ $(B)$, breaks inversion symmetry and opens a trivial band gap $\Delta_m = 2m$. The third term is a complex second-neighbor hopping that breaks time-reversal symmetry and opens a topological gap $\Delta_\mathrm{H} = 6\sqrt{3}t_2$ (here we set $\phi_{ij}=\pi/2$). When both gap-opening terms are nonzero, the total band gap is $\Delta = |\Delta_m - \Delta_{\mathrm{H}}|$. When $\Delta_{m} > \Delta_{\mathrm{H}}$ the system is a trivial insulator, otherwise it is a Chern insulator with $\mathcal{C}=1$ \cite{Thouless1982Aug, Haldane1988Oct}. This model well describes the class of Chern insulators and the related quantum anomalous Hall phase \cite{RevModPhys.95.011002, 10.1063/5.0100989}.

We now consider the three primary phases that emerge in this model. In the trivial phase, equivalent to gapped graphene, we let $\Delta_m=2$ and $\Delta_{\mathrm{H}}=0$. In the topological phase with inversion symmetry, $\Delta_m=0$ and $\Delta_{\mathrm{H}}=2$. Finally, in the topological phase with broken inversion symmetry, $\Delta_m=2$ and $\Delta_{\mathrm{H}}=4$. This last phase has been shown to behave differently in the presence of disorder than the inversion-symmetric one \cite{JoseHugoKPM}. The band structures of these phases are shown in the left panels of Fig.\ \ref{fig:Haldanechemea}. In all cases the total band gap is $\Delta = 2$.

\begin{figure}[tbh]
\includegraphics[width=1\columnwidth]{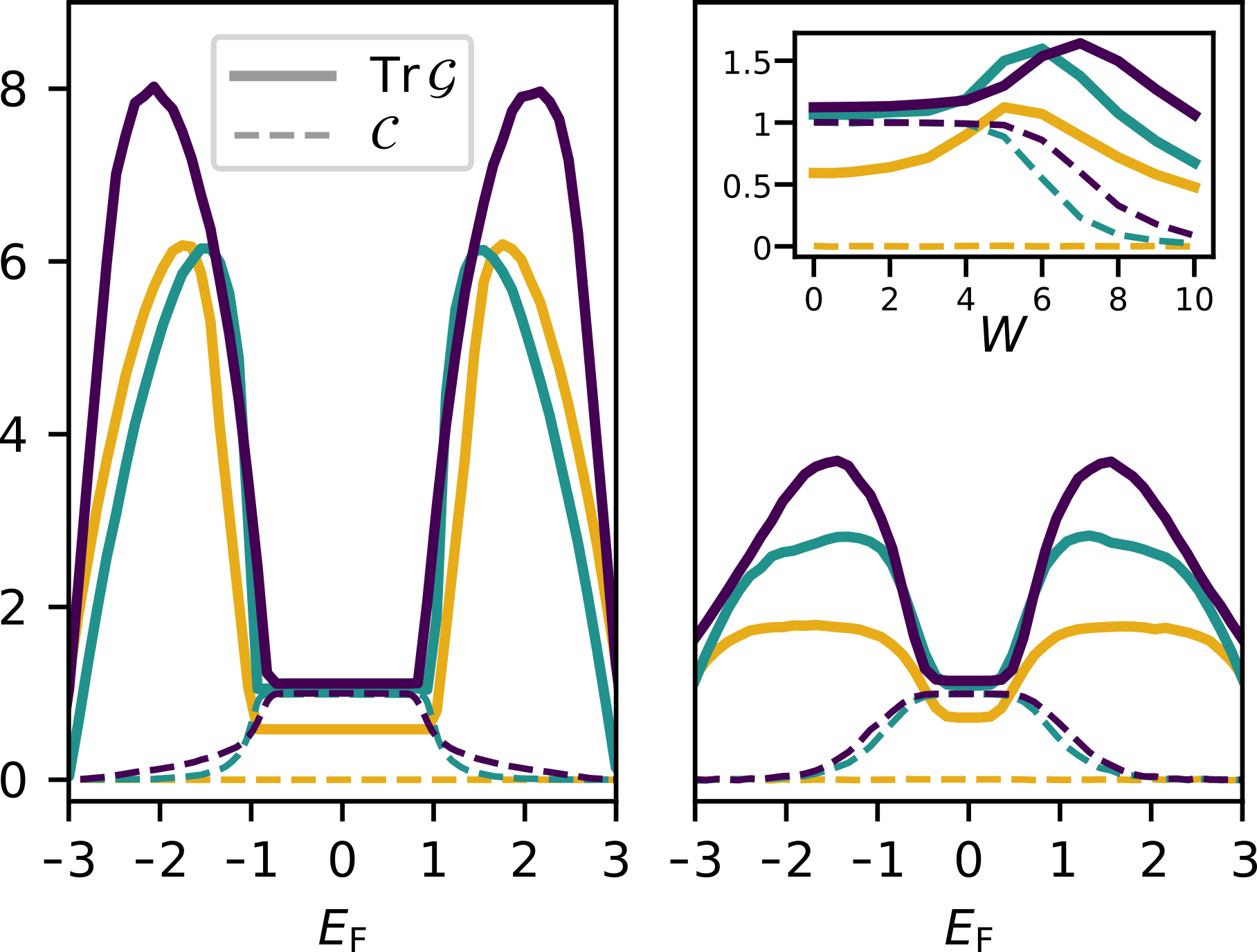}
\caption{\label{fig:Andersondisorder}Left panel: IQM (solid lines) and Chern number (dashed lines) vs.\ Fermi energy without disorder. The trivial, purely topological ($m=0$), and inversion-symmetry-broken topological ($m=1$) phases are shown in yellow, green, and purple. Right panel: IQM and Chern number vs.\ Fermi energy for Anderson disorder strength $W=3$. Inset: IQM and Chern number vs.\ disorder strength at $E_{\mathrm{F}}=0$.}
\end{figure}

Figure \ref{fig:Andersondisorder} (left panel) shows the Chern number, $\mathcal{C}$ (dashed lines), and the trace of the IQM, $\Tr\mathcal{G}$ (solid lines), for the three phases without disorder. As expected, $\mathcal{C} = 1$ inside the gap of the topological phases, and is zero for the trivial phase. Meanwhile, broken inversion symmetry extends the topological phase ($\mathcal{C} > 0$) to higher energies. In this phase the $K_\pm$ valleys have different gaps, $\Delta = 2$ and $6$ (bottom left panel of Fig.\ \ref{fig:Haldanechemea}). For $E_\mathrm{F} \in [-3, -1]\cup[1, 3]$, only the $K_+$ valley contributes and thus $\mathcal{C}$ is finite for a wider energy range than for the $m=0$ topological phase. On the other hand, the IQM is nonzero both inside and outside the gap, and is bounded from below by $\mathcal{C}$ in all cases. In the gap, the IQM is reduced in the trivial phase compared to the topological phases, indicating a higher degree of localization.
Outside the gap, the system is metallic and the wave functions are delocalized, as a scaling analysis reveals a divergent IQM in the thermodynamic limit \cite{insulatingtheoryofmatter, suppmaterial}. Minor particle-hole asymmetries arise from numerical noise inherent in the stochastic approximation of the trace, but these are progressively reduced by greater statistical averaging \cite{suppmaterial}. 

We next consider the IQGT in the presence of Anderson disorder \cite{Anderson1958Mar}, modeled as an onsite potential,

\begin{equation}
    \hat{H}_W = \sum_i \epsilon_i c_i^{\dagger} c_i,
\end{equation}

where $\epsilon_i$ are randomly distributed in the interval $\left[-W/2,W/2\right]$. In two dimensions, such disorder induces a metal-insulator transition for which all states are localized at all energies \cite{ReviewLocalization}. Figure \ref{fig:Andersondisorder} (right panel) shows $\mathcal{C}$ and the IQM for disorder strength $W=3$. Here the gap is shrunk, owing to disorder-induced broadening and the introduction of in-gap localized states. Outside the gap, the IQM is suppressed in all cases. A scaling analysis shows that these states that were metallic without disorder are now localized in the limit $M,A\rightarrow\infty$ \cite{suppmaterial}, consistent with the metal-insulator transition.

\begin{figure*}[tbh]
\includegraphics[width=1.5\columnwidth]{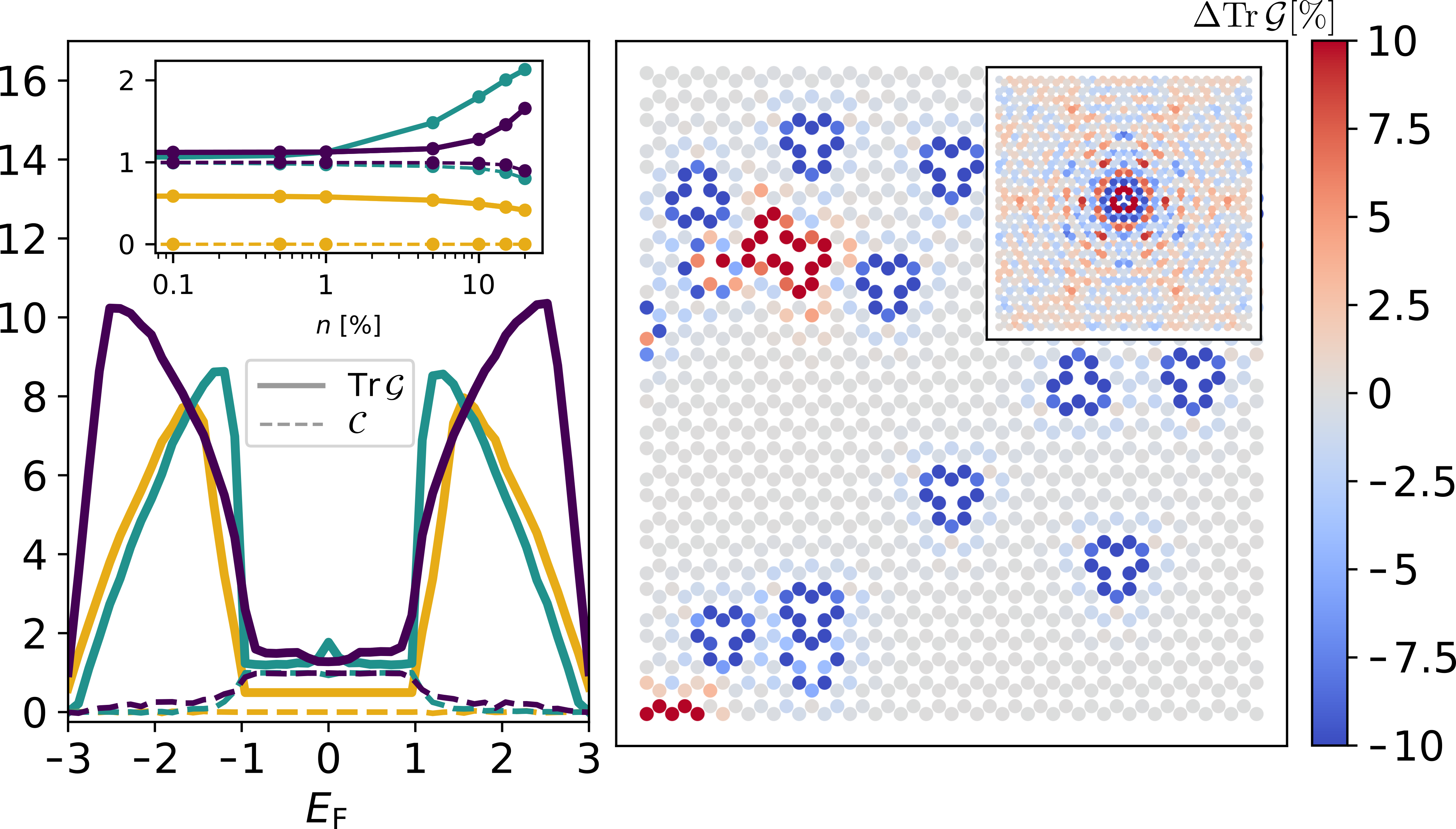}
\caption{\label{fig:figurevacancies}Left panel: IQM (solid lines) and Chern number (dashed lines) vs.\ Fermi energy for a vacancy concentration of $n=10\%$. The trivial, purely topological ($m=0$), and inversion-symmetry-broken topological ($m=1$) phases are shown in yellow, green, and purple. Left inset: IQM and Chern number for $n \in [0.1,20\%]$ at $E_{\mathrm{F}}=0$. Right panel: real-space fluctuations of the IQM, $\Delta\mathrm{Tr}\mathcal{G} = 100\% \cdot (\mathrm{Tr}\mathcal{G} - \mathrm{Tr}\mathcal{G}_{\mathrm{clean}}) / \mathrm{Tr}\mathcal{G}_\mathrm{clean}$, for $n=1\%$ of vacancies in the topological phase with $m=0$ at $E_{\mathrm{F}}=0$. Inset: real-space fluctuations of the IQM around a single impurity at high energy, $E_{\mathrm{F}} = 2$.}
\end{figure*}

The inset of Fig.\ \ref{fig:Andersondisorder} shows $\mathcal{C}$ and the IQM at mid-gap ($E_{\mathrm{F}}=0$) for varying disorder strength. Both quantities are robust for a wide range of $W$ and only decay for disorder strength larger than the gap ($W \gtrsim 4$). The topological phase with broken inversion symmetry exhibits the slowest decay as well as the highest overall values. These features may be linked to the absence of intervalley scattering in this phase, thus reducing localization.

We also see that for $W \lesssim 4$ the IQM actually increases with increasing disorder. This may be attributed to the mixing of states into the gap via disorder-induced broadening. For small $W$, the localization length of these states is expected to be longer than the in-gap states, effectively increasing the IQM. At higher $W$, Anderson localization then fully takes over. A similar effect has also been observed in twisted bilayer graphene at magic angle \cite{PedroGuerrero2024Jan}. 

Finally, we explore the impact of vacancy defects, shown schematically in Fig.\ \ref{fig:Haldanechemea} (bottom right). Vacancies are modeled by randomly removing $n \cdot N$ lattice sites, with $n$ the vacancy concentration and $N$ the number of sites in the clean system. Their impact on quantum transport has been well scrutinized in Dirac materials, with the presence of zero energy modes leading to nontrivial transport phenomena \cite{PhysRevLett.110.196601, PhysRevLett.115.106601, marsal2024enhanced}.

Figure \ref{fig:figurevacancies} (left panel) shows the IQM and the Chern number for a vacancy concentration of $n=10\%$, and the inset shows them at $E_{\mathrm{F}}=0$ for $n \in [0.1,20\%]$. Outside the gap the IQM becomes localized, similar to the case of Anderson disorder in Fig.\ \ref{fig:Andersondisorder}. In the gap (see inset), different behaviors develop depending on the phase. In the trivial phase, the IQM is weakly reduced with increasing vacancy concentration, while in the topological phases the IQM first slightly decreases and then eventually increases for large concentration, with this increase sharper without broken inversion symmetry ($m=0$).

To understand this, in Fig.\ \ref{fig:figurevacancies} (right panel) we consider the real-space distribution of the IQM of the topological phase ($m=0$) at $E_\mathrm{F} = 0$ for $n=1\%$ (see the SM for corresponding densities of states (DOS) \cite{suppmaterial}). In this phase, vacancies locally reduce the IQM and introduce in-gap impurity states which arise from the bulk-edge correspondence \cite{Vacacancies, boundstatesinvacancies}. At high vacancy concentration, the real-space projection enables us to understand the increase in the IQM, as neighboring impurity states are more likely to overlap (dark red region), yielding an enhancement of delocalization consistent with the increase of the IQM in the left panel. The real-space projection at $n=10\%$ is shown in the SM \cite{suppmaterial}.

In the topological phase with broken inversion symmetry ($m=1$), vacancies create two peaks in the DOS at $E_{\mathrm{F}} = \pm 0.31$ \cite{suppmaterial}, as the $A/B$ sites are no longer equivalent.
Like the Anderson disorder case, this phase is generally less localized than the pure topological phase. This weaker dependence on disorder also occurs at the energies of the impurity peaks \cite{suppmaterial}. As each peak is localized on one sublattice, this may be connected to prior studies in graphene and topological insulators indicating that vacancies or hydrogen atoms distributed on one sublattice have a weaker impact than those distributed on both \cite{ROCHE20121404, Leconte2011, PhysRevLett.107.016602, boundstatesinvacancies, marsal2024enhanced}. Finally, the IQM in the trivial phase is only weakly reduced, as in-gap states do not form owing to a lack of edge states \cite{boundstatesinvacancies}. The real-space projection at high energy (Fig.\ \ref{fig:figurevacancies}, right inset) reveals long-range patterns characteristic of Friedel oscillations \cite{FriedeloscillationsGraphene, OscillationsFluorRoche, FriedelOscillationsinFL}. This is present in all phases \cite{suppmaterial}, but further study is needed to correlate such behavior with the impact on transport.

\section{Summary and outlook}
We have introduced an efficient linear-scaling algorithm to compute the integrated quantum geometric tensor of large inhomogeneous systems. We illustrated the method by exploring the IQGT in trivial and topological systems with Anderson disorder and vacancies. The integrated quantum metric was verified to be lower bounded by the Chern number, and its scaling with disorder provides information about localization-delocalization transitions, which depend on the type and strength of the disorder, as well as the topological phase.

We focused on a well-studied model to make meaningful comparisons between the integrated quantum metric and known transport properties, but the method is applicable to any real-space Hamiltonian. Future work could also explore other materials such as topological insulators and semimetals, or Moiré systems, in which disorder effects might be key in understanding their emergent properties. Similarly, aperiodic systems such as quasicrystals \cite{PhysRevLett.132.086402, PhysRevB.108.115109, Shimasaki2024} could be more efficiently studied within this new approach. Finally, interaction effects may also be included, provided they can be condensed into a single-particle Hamiltonian. For example, recent work uses the KPM to efficiently derive a self-consistent mean-field representation of twisted bilayer graphene, among other systems \cite{Lado2024}, which may then be easily incorporated into our methodology for calculating the integrated quantum geometric tensor. Such mean field effects have also proven significant in the study of the quantum metric \cite{FirstPavi, AnotherPavi}.

\begin{acknowledgments}
J.M.R.\ acknowledges P.A.\ Guerrero and L.M.\ Canonico for fruitful discussions.
We acknowledge funding from MCIN/AEI /10.13039/501100011033 and European Union "NextGenerationEU/PRTR” under grant PCI2021-122035-2A-2a. ICN2 is funded by the CERCA Programme/Generalitat de Catalunya and supported by the Severo Ochoa Centres of Excellence programme, Grant CEX2021-001214-S, funded by MCIN/AEI/10.13039.501100011033. This work is also supported by MICIN with European funds‐NextGenerationEU (PRTR‐C17.I1) and by and 2021 SGR 00997, funded by Generalitat de Catalunya.

\end{acknowledgments}

\end{document}